\documentclass{article}
\usepackage{amssymb}
\usepackage{epsfig}
\usepackage{color}
\usepackage{amsfonts}
\usepackage{amsmath}
\usepackage{bbold}
\usepackage{graphicx}
\usepackage{mathtools}
\usepackage{tikz}

\reversemarginpar
\newcommand{\spacer}{\rule[0cm]{0cm}{0cm}}

\newtheorem{theorem}{Theorem}
\newtheorem{lemma}{Lemma}
\newtheorem{proposition}{Proposition}
\newtheorem{corollary}{Corollary}

\newcounter{statementnumber}
\renewcommand{\thestatementnumber}{\arabic{section}.\arabic{statementnumber}}

\newenvironment{proof}{\par\noindent\textsc{Proof.}}{\nopagebreak\spacer\hfill $\square$}

\newcounter{figurecount}
\def\mb{$\begin{displaystyle}}
\def\me{\end{displaystyle}$\ }

\def\Id{\mbox{\rm Id}}
\DeclareMathOperator{\wt}{wt}

\DeclareMathOperator{\tr}{tr}

\newcommand{\ket}[1]{\left| #1 \right\rangle}
\newcommand{\bra}[1]{\left\langle #1 \right|}

\newcommand{\twotwo}[4]{\left[ \begin{array}{cc} #1 & #2 \\
                        #3 & #4 \end{array} \right]}
\newcommand{\R}{{\mathbb R}}

\newcommand{\C}{{\mathbb C}}

\DeclareMathOperator{\Herm}{Herm}
\newcommand{\purestatespace}[1]{{\mathcal H}{_{#1}}}
\newcommand{\purewernerspace}[1]{\purestatespace{#1}^{\mathcal W}}
\newcommand{\matrixspace}[1]{{\mathcal L}(\purestatespace{#1})}
\newcommand{\matrixwernerspace}[1]{{\mathcal L}(\purestatespace{#1})^{\mathcal W}}
\newcommand{\hermitianspace}[1]{{\Herm}(\purestatespace{#1})}
\newcommand{\hermitianwernerspace}[1]{{\Herm}(\purestatespace{#1})^{\mathcal
    W}}

\newcommand{\ncc}{\rm NCC} % NonCrossing Chord diagrams
\newcommand{\pizzastate}{\ket{P}}
\newcommand{\Pizza}{m(\pizzastate)}
\newcommand{\pb}{\scalebox{.5}{\begin{picture}(0,0)%
\includegraphics{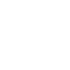}%
\end{picture}%
\setlength{\unitlength}{3947sp}%
\begingroup\makeatletter\ifx\SetFigFont\undefined%
\gdef\SetFigFont#1#2#3#4#5{%
  \reset@font\fontsize{#1}{#2pt}%
  \fontfamily{#3}\fontseries{#4}\fontshape{#5}%
  \selectfont}%
\fi\endgroup%
\begin{picture}(324,324)(2839,-523)
\put(3001,-436){\makebox(0,0)[b]{\smash{{\SetFigFont{17}{20.4}{\rmdefault}{\mddefault}{\updefault}{\color[rgb]{0,0,0}$+$}%
}}}}
\end{picture}%
}}
\newcommand{\nb}{\scalebox{.5}{\begin{picture}(0,0)%
\includegraphics{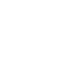}%
\end{picture}%
\setlength{\unitlength}{3947sp}%
\begingroup\makeatletter\ifx\SetFigFont\undefined%
\gdef\SetFigFont#1#2#3#4#5{%
  \reset@font\fontsize{#1}{#2pt}%
  \fontfamily{#3}\fontseries{#4}\fontshape{#5}%
  \selectfont}%
\fi\endgroup%
\begin{picture}(324,324)(2839,-523)
\put(3001,-436){\makebox(0,0)[b]{\smash{{\SetFigFont{17}{20.4}{\rmdefault}{\mddefault}{\updefault}{\color[rgb]{0,0,0}$-$}%
}}}}
\end{picture}%
}}
\newcommand{\zb}{\scalebox{.5}{\begin{picture}(0,0)%
\includegraphics{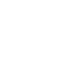}%
\end{picture}%
\setlength{\unitlength}{3947sp}%
\begingroup\makeatletter\ifx\SetFigFont\undefined%
\gdef\SetFigFont#1#2#3#4#5{%
  \reset@font\fontsize{#1}{#2pt}%
  \fontfamily{#3}\fontseries{#4}\fontshape{#5}%
  \selectfont}%
\fi\endgroup%
\begin{picture}(324,324)(2839,-523)
\put(3001,-436){\makebox(0,0)[b]{\smash{{\SetFigFont{17}{20.4}{\rmdefault}{\mddefault}{\updefault}{\color[rgb]{0,0,0}$0$}%
}}}}
\end{picture}%
}}
\title{Werner states from diagrams}
\author{David W. Lyons\footnote{{\tt lyons@lvc.edu}}, Cristina Mullican, Adam Rilatt, Jack D. Putnam\\
{\small Mathematical Sciences, Lebanon Valley College, Annville,
    PA, USA}}
\date{revised: 7 April 2023}

\begin{document}

\maketitle

\begin{abstract}
We present two results on multiqubit Werner states, defined to be those
states that are invariant under the {collective} action of any given single-qubit
unitary that acts simultaneously on all the qubits.  Motivated by the
desire to characterize entanglement properties of Werner states, we
construct a basis for the real linear vector space of Werner invariant
Hermitian operators on the Hilbert space of pure states; it follows
that any mixed Werner state can be written as a mixture of
these basis operators with unique coefficients. Continuing a study of
``polygon diagram'' Werner states constructed in earlier work, with a
goal to connect diagrams to entanglement properties, we consider a
family of multiqubit states that generalize the singlet, and show that
their 2-qubit reduced density matrices are separable.
\end{abstract}

\setcounter{equation}{0}
\setcounter{statementnumber}{0}

\section{Introduction}\label{introsection}

Motivated by practical applications in computation, cryptography, and
metrology, quantum information theory has been instrumental in shedding
light on fundamental theoretical questions in physics and computer
science. This includes violation of Bell inequalities and local hidden
variable theories~\cite{bell64,werner89,greenberger89}, new proofs of
classical information theorems~\cite{druckerdewolf2011}, and new
physical principles such as information
causality~\cite{pawloskietalinfcausality2009}.

Certain classes of states have played significant roles in theoretical
and applied developments in quantum information. This paper focuses on
multiqubit Werner states, defined by their invariance under
the action of local unitaries of the form $U^{\otimes n}$, for all
1-qubit unitaries $U$, and $n$ is the number of qubits.
Originally introduced in 1989 for two particles to
distinguish between classical correlation and Bell inequality
satisfaction~\cite{werner89}, Werner states have been used in the description of noisy
quantum channels~\cite{lee00}, as examples in nonadditivity
claims~\cite{shor01}, for hiding classical data in quantum
states~\cite{eggeling2002hiding}, in the study of deterministic
purification~\cite{short09}, and for coding in a way that protects
against the loss of a qubit~\cite{migdal2011}.

Significant results on the properties of Werner states include detailed
understanding of structure and entanglement properties for bipartite and
tripartite systems of arbitrary local
dimension~\cite{werner89,eggelingwernerPhysRevA.63.042111} and general
results on entanglement
witnesses~\cite{huber2021positive,balanzo2021positive}. In this paper,
we extend our own previous work on pure and mixed Werner states of
arbitrarily many qubits. In~\cite{su2blockstates}, we construct linear
bases for the Hilbert spaces of pure Werner states, parameterized by
combinatorial objects called chord diagrams.
%% An appealing feature of the 
%% analysis is that entanglement properties of a superposition of these
%% basis states can be read directly from the associated
%% diagrams.
In~\cite{wernerstructure}, we construct mixed Werner states
from another type of diagram called polygon diagrams that are directly
related to properties of separability and cyclic permutational symmetry.

This paper builds on our diagram-based analyses towards further
structural understanding of mixed Werner states. In
Section~\ref{nccbasissection}, we use our chord diagram basis for
$2n$-qubit pure Werner states to construct a basis for the real vector
space of Werner invariant Hermitian operators on the Hilbert space for
$n$ qubits; it follows that any $n$-qubit mixed Werner state can be
written as a mixture of these basis operators with unique
coefficients. Motivation for this construction comes from the success of
Werner and Eggeling's precise mapping of separability regions in the
space of coefficients with respect to a specific basis for tripartite
Werner states~\cite{eggelingwernerPhysRevA.63.042111}.  Towards the goal
of further connecting polygon diagrams to entanglement properties of
states constructed from them, we consider a family of polygon diagram
states that generalize the singlet to many qubits, and show that their
2-qubit reduced density matrices are separable
in~Section~\ref{rdmentsect}. This result can be viewed as a case study
related to recent work of Bernards and
G\"uhne~\cite{bernards2022multiparticle} where they show, in their study
of absolutely maximally entangled (AME) states, that 2-party reduced
density matrices of pure Werner states are never maximally mixed. In a
distributed quantum computation scenario, these polygon Werner states
provide a multipartite entanglement resource that does not allow
2-party shared entanglement, thus affording some protection against
dishonest pairs of parties.

We begin with preliminary facts and notation in
Section~\ref{prelimsect}. We give a self-contained account of our
construction of pure Werner states from chord diagrams, and another
construction of a family of mixed Werner states that generalize the
singlet state, in
Section~\ref{diagramconstrsection}. Some proofs involving longer or more
technical derivations are given in the appendix.

\section{Preliminaries}\label{prelimsect}

An $m$-qubit pure state $\ket{\psi}$ is Werner invariant if $U^{\otimes
  m}\ket{\psi}\propto \ket{\psi}$ for all 1-qubit unitary operators
$U$. An $m$-qubit mixed state $\rho$ is Werner invariant if $U^{\otimes
  m} \rho (U^\dagger)^{\otimes m}=\rho$ for all 1-qubit unitary operators
$U$. More generally, an operator $A$ on $m$-qubit states is Werner
invariant if $U^{\otimes m} A (U^\dagger)^{\otimes m}=A$ for all 1-qubit
unitary operators $U$.

We will write $\purestatespace{m},\matrixspace{m},\hermitianspace{m}$ to denote the
Hilbert space of pure states, the space of operators on Hilbert space, and
the space of Hermitian operators on Hilbert space, respectively. 
We will write
$\purewernerspace{m},\matrixwernerspace{m},\hermitianwernerspace{m}$
to denote the corresponding Werner invariant subspaces. In these
notations, the set of mixed states of $m$ qubits is a subset of
$\hermitianspace{m}$, and the $m$-qubit Werner invariant mixed states
are a subset of $\hermitianwernerspace{m}$.

We write $Z,X$ denote the 1-qubit Pauli operators
$Z=\twotwo{1}{0}{0}{-1}$, $X=\twotwo{0}{1}{1}{0}$ with respect to a
given computational basis, and we use the notation $A^{(k)}$, where $A$
is either $Z$ or $X$, to denote the 1-qubit operator $A$ acting on the
$k$-th qubit of a multiqubit state, i.e., $A^{(k)}$ is the operator
$I^{\otimes (k-1)}\otimes A\otimes I^{(m-k)}$ in $\matrixspace{m}$.

{We will use the following formulas for establishing Werner invariance in
sections below.} An $m$-qubit pure state $\ket{\psi}$ is Werner invariant
if the following two equations hold
  \begin{align}
    \left(\sum_k Z^{(k)}\right)\cdot \ket{\psi} &= 0\label{Acondpure}\\
    \left(\sum_k X^{(k)}\right)\cdot \ket{\psi} &= 0\label{Ccondpure}
  \end{align}
and an $m$-qubit mixed state $\rho$ is Werner invariant if the following
two equations hold.
  \begin{align}
    \left[\left(\sum_k Z^{(k)}\right),\rho\right] &= 0\label{Acondmixed}\\
    \left[\left(\sum_k X^{(k)}\right),\rho\right] &= 0\label{Ccondmixed}
  \end{align}
{While these criteria for Werner invariance are well-known, we provide a
proof in the appendix for the sake of self-containedness.}

We will use the following notation for bit strings. Given an $m$-bit
string $I=i_1i_2\ldots i_m$, we write $\wt I$ to denote the Hamming
weight $\wt I = \sum_k i_k$. We write $i_k^c$ to denote the complement
$i_k + 1 \pmod{2}$ of the $k$th bit $i_k$, and we write $I_\ell$ to denote
the string $i_1i_2 \ldots i_{\ell -1} i_\ell^c  i_{\ell + 1} \ldots i_m$, that is, the
string $I$ with only the $\ell$-th bit complemented, and the other bits left
unchanged. We write $I^c$ to denote the string $i_1^ci_2^c\ldots i_m^c$.

Given a bit string $J=j_1,j_2,\ldots,j_d$, we write $J^k$ to denote the $kd$-bit
string obtained by concatenating $J$ with itself $k$ times. For example,
$$ (011)^3 = 011011011.
$$
A bit string $I$ is called {\em periodic} if $I=J^k$ for some $k>1$, and
is called {\em aperiodic} otherwise.

\section{Pure and Mixed Werner State Constructions from Diagrams}\label{diagramconstrsection}

This background section provides details from previous work that is
needed for the new results in the sections that follow.

It is straightforward to check that the singlet state
$\ket{s}=\frac{1}{\sqrt{2}}\left(\ket{01}-\ket{10}\right)$ is Werner
invariant. It follows that any product of singlet states is also Werner
invariant. Not obvious, but true nonetheless, is that any pure Werner
invariant state must be a superposition of products of
singlets~\cite{su2blockstates}. Thus, to describe pure Werner states, it
is natural to make use of {\em chord diagrams} to keep track of which
pairs of qubits are entangled in a product of singlets. A chord diagram
with $2n$ nodes is a partition of the set $\{1,2,\ldots,2n\}$ into
two-element subsets, called chords. The diagram is drawn with points
labeled $1,2,\ldots,2n$ consecutively around a circle, with a line
segment connecting each pair $\{a,b\}$ in the chosen partition. The
figures in the left column of Table~\ref{tab:n3eg} show examples.

For our basis construction in the next section, it will be convenient to consider {\em oriented}
chord diagrams, where ordered pairs are used to denote chords, instead of two-element
sets. We will write $(a,b)$ do denote the directed chord starting at
vertex $a$ and ending at vertex $b$, and we write $\ket{s}_{a,b}$ to denote
the singlet
$$\ket{s}_{a,b} = \frac{1}{\sqrt{2}}\left(\ket{0}_a \ket{1}_b - \ket{1}_a \ket{0}_b\right)
$$
associated to the directed chord $(a,b)$.
Given an oriented chord diagram ${\mathcal  D} = \{(a_k,b_k)\}_{1\leq
  k\leq n}$, we define the state $\ket{\mathcal D}$ to be the product of singlets
$$ \ket{{\mathcal D}}=\bigotimes_{1\leq k\leq n} \ket{s}_{a_k,b_k}.
$$
In the Hilbert space $(\C^2)^{\otimes 2n}$ of the composite
system of $2n$ qubits in order $\{1,2,\ldots,2n\}$, the orientation
reversal of a chord flips the sign of a diagram state. That is, if
${\mathcal D},{\mathcal D}'$ share all but one of the same oriented
chords, but $(a,b)$ is a chord in ${\mathcal D}$ and $(b,a)$ is a chord
in ${\mathcal D}'$, we have $\ket{\mathcal D}=-\ket{\mathcal D}'$.
The coefficients $c_K$ in the expression $\ket{\mathcal D}=\sum_K c_K
\ket{K}$ for $\ket{\mathcal D}$ in the
computational basis parameterized by $2n$-bit strings $K=k_1k_2\ldots
k_{2n}$ are given by 
\begin{equation}\label{diagramstatecoeff}
  c_K = 
  \begin{cases}
    \prod_{\ell = 1}^n (-1)^{k_{a_\ell}} & \text{if
      $k_{a_\ell}=k_{b_l}^c$ for $1\leq \ell\leq n$}\\
    0 & \text{otherwise}
  \end{cases}.
\end{equation}

A chord diagram is said to be {\em noncrossing} if there are no
intersections of chords in the geometric picture. We will write $\ncc$
to denote the set of all $2n$-node noncrossing chord diagrams, where $n$
will be understood from context, where each chord $\{a,b\}$ with $a<b$
has the orientation $(a,b)$. It is a remarkable
fact~\cite{su2blockstates} the the set of singlet products
corresponding to $\ncc$ form a $\C$-linear basis for pure Werner
states. Table~\ref{tab:n3eg} shows the 5 noncrossing chord diagrams for 6 qubits.

\begin{figure}
    \begin{center}
      \scalebox{.4}{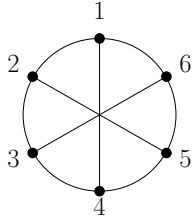}  
    \end{center}
    \caption{The ``pizza'' diagram ${\mathcal P}_0$ for $2n=6$
      qubits. The unnormalized pizza state is $\ket{P}=
      \ket{000111}-\ket{111000}-\ket{001110}+\ket{110001}-\ket{010101}+\ket{101010}+\ket{011100}-\ket{100011}$.}
    \label{tab:pizzafig}
  \end{figure}

We will use the following singlet product state in our constructions for
mixed Werner states in the next section. We define the ``pizza diagram''
${\mathcal P}_0$ to be the chord diagram ${\mathcal P}_0 =
\{(i,i+n) \colon 1\leq i\leq n\}$. See Figure~\ref{tab:pizzafig}. For convenience, we
rescale the state $\ket{{\mathcal P}_0}$ to define the (unnormalized)
$2n$-qubit, Werner invariant ``pizza state''
$\ket{P}=2^{n/2}\ket{{\mathcal P}_0}$.  Using notation from
Section~\ref{prelimsect} above, the pizza state can be expressed
as follows.
\begin{equation}
  \label{pizzastatecompbasis}
  \pizzastate = \sum_I (-1)^{\wt I} \ket{I}\ket{I^c} =
  2^{n/2}\ket{{\mathcal P}_0}
\end{equation}

Next, we construct a family of mixed Werner states $\rho_m$ that
generalize the density matrix $\rho_2=\ket{s}\bra{s}$ of the
singlet.\footnote{The states $\rho_m$ appear as tensor factors in a diagrammatic
construction for mixed Werner states that generalizes the chord diagram
construction for pure Werner states. The results of Section~\ref{rdmentsect} do
not require the full polygon state construction, so we limit our
discussion to only the necessary details for $\rho_m$.}
The first step is to construct a pure state $C(I)$ for every aperiodic
$m$-bit string $I=i_1i_2\cdots i_m$, given by
\begin{equation}\label{cyclicistatedef}
   C(I) = \frac{1}{\sqrt{m}} \sum_{k=0}^{m-1} \omega^{k} \ket{\pi^k I}
\end{equation}
where $\omega = e^{2\pi i/m}$ and $\pi$ is the cyclic permutation of
$\{1,2,\ldots,m\}$ given by $j\to j-1 \pmod{m}$. For example, we have
$$C(001) = \frac{1}{\sqrt{3}}(\ket{001} + e^{\frac{2\pi i}{3}}\ket{010} +
  e^{\frac{4\pi i}{3}}\ket{100}).
  $$
  (Note that if $I$ is periodic, then $C(I)=0$, and is therefore not a
  state.) Now we define $\rho_m$ by
  \begin{equation}\label{wernergondef}
     \rho_m = \frac{1}{A(m)} \sum_{\text{aperiodic }I} C(I)C(I)^\dagger
  \end{equation}
  where $A(m)$ is the number of aperiodic $m$-bit strings~\cite{oeisA027375}.
It is easy to check that $\rho_2=\ket{s}\bra{s}$ is the density matrix
for the singlet state, and it is a fact~\cite{wernerstructure} that $\rho_m$
is Werner invariant for all $m\geq 1$. 

%% Next, we generalize the chord diagram construction of pure Werner states
%% to a polygon diagram construction of mixed Werner states. A
%% polygon diagram with $n$ nodes is a partition of the set $\{1,2,\ldots,
%% n\}$ into subsets called polygons, with no restriction on the size of
%% the partition subsets. The polygon diagram is drawn with points labeled
%% $\{1,2,\ldots,n\}$ consecutively around a circle, with a convex $m$-gon
%% drawn for each partition set $\{a_1,a_2,\ldots,a_m\}$. See Figure
%% blank. Just as we did for chord diagrams, we will construct a Werner
%% state for each polygon in the partition, then we define the state
%% associated to the polygon diagram to be the tensor product of the states
%% associated to each partition set (cite ourselves).

\section{A mixed Werner basis construction}\label{nccbasissection}

In this section we construct a basis for the real vector space
$\hermitianwernerspace{n}$.  It follows that any $n$-qubit mixed Werner
state can be written uniquely as an $\R$-linear combination of matrices
in this set. The overall strategy is to map a known basis for $2n$-qubit
pure Werner states (the noncrossing chord diagram states) to a basis of
operators on the Werner invariant subspace of Hermitian operators
$n$-qubit state space.

To begin, let ${\cal H}=\purestatespace{2n}=(\C^2)^{\otimes n}\otimes (\C^2)^{\otimes n}$ denote the Hilbert space of
$2n$-qubit states with computational basis $\{\ket{I}\ket{J}\}$, where $I,J$
each range over the set of all $n$-bit strings. Let $(\purestatespace{2n})_{\R}$
denote the subspace of states with real coefficients in the
computational basis.
$$ (\purestatespace{2n})_{\R} = \left\{\sum_{IJ} c_{IJ} \ket{I}\ket{J}\colon c_{IJ} \in \R\right\}
$$ Let $m$ denote the $\R$-linear map $m\colon
(\purestatespace{2n})_{\R} \to \matrixspace{n}$ that takes the
computational basis vector $\ket{I}\ket{J}$ to $\ket{I}\bra{J}$ (note
that $m$ is {\em not} $\C$-linear). Observe that $\ket{\mathcal D}$ lies
in $(\purestatespace{2n})_{\R}$ for any chord diagram (crossing or
noncrossing), so that $m(\ket{\mathcal D})$ is defined.

Next, we establish useful properties of the ``pizza operator''
$m(\pizzastate)$ obtained by applying $m$ to the (unnormalized) pizza
state $\ket{P}$. The symbol $Y$ denotes
the Pauli operator $Y=\twotwo{0}{-i}{i}{0}$.

\begin{proposition}\label{pizzaoperatorproplemma}{\bf (Properties of the pizza operator)}
    The following hold. 
    \begin{enumerate}
    \item[(i)] $\Pizza = \sum_I (-1)^{\wt I} \ket{I}\bra{I^c}$      
    \item[(ii)] $\Pizza^T = (-1)^n \Pizza$
    \item[(iii)] $\Pizza = (iY)^{\otimes n} = 2^{n/2}m(\ket{s})^{\otimes n}$
    \item[(iv)] $\Pizza^2 = (-1)^n \Id$
    \end{enumerate}
    \end{proposition}

\begin{proof}
  For~(i), apply the definition of $m$ to the expression~(\ref{pizzastatecompbasis}) for
  the pizza state. For~(ii), use $\wt I+\wt I^c=n$, so $(-1)^{\wt I^c} =
  (-1)^n(-1)^{\wt I}$. Checking~(iii) is a straightforward computation,
  and~(iv) follows from $(iY)^2 = -\Id$. 
\end{proof}

The next Proposition establishes key properties of products $\Pizza
m(\ket{\mathcal D})$.

\begin{proposition}\label{pizzatimesdiagramprop} Let ${\mathcal D}$
  be any chord diagram, crossing or noncrossing. The following hold.
  \begin{enumerate}
  \item[(i)]   $\Pizza m(\ket{\mathcal D})= m(\ket{\mathcal D})\Pizza $
    \item[(ii)] $\Pizza m(\ket{{\mathcal D}})$ is Werner invariant
  \end{enumerate}
\end{proposition}

\begin{proof} The proof of~(i) requires only simple observations about
  products of singlets. The proof of~(ii) uses Werner invariance
  criteria~(\ref{Acondpure})--(\ref{Ccondmixed}) to verify that $\Pizza
  m(\ket{\mathcal D})$ is Werner invariant. Details are given in the
  appendix.
\end{proof}

In the remainder of this section, we describe how to construct a basis
for $\hermitianwernerspace{n}$ from the Werner invariant matrices in
part~(ii) of Proposition~\ref{pizzatimesdiagramprop} above. For
compactness and readability, let
$$A_{\mathcal D}:=\Pizza m(\ket{{\mathcal D}}).$$
We will show that there
are two possibilities for $A_{\mathcal D}$, depending on whether
${\mathcal D}$ has half-turn rotational symmetry.
For ${\mathcal D}=\{(a_k,b_k)\colon 1\leq k\leq n\}$, we define $R_{180}{\mathcal D}$ by
$$R_{180}{\mathcal
  D}=\{(a_k+n,b_k+n)\colon 1\leq k\leq n\}$$
where addition in the last expression is
taken mod $2n$.
To say that a diagram ${\mathcal D}$ has half-turn rotational symmetry
means that ${\mathcal D}$,$R_{180}{\mathcal D}$ are equal as unoriented chord
diagrams, which is the same as $\ket{\mathcal D} = \pm\ket{R_{180}{\mathcal
    D}}$. We will show that 
      \begin{itemize}
      \item $A_{\mathcal D}$ is symmetric if ${\mathcal D}$ has
        half-turn symmetry, and
        \item $A_{\mathcal D},A_{R_{180}{\mathcal D}}$
      are distinct and are transposes of one other (up to a sign) if ${\mathcal D}$
      does not have half-turn symmetry.
      \end{itemize}
We then construct a set of Hermitian matrices from linear combinations
of the $A_{\mathcal D}$, and finally, we argue why this set forms a basis.
%% that consists of all the
%% $A_{\mathcal D}$ for ${\mathcal D}$ with half-turn symmetry, and of
%% pairs $(A_{\mathcal D} + A_{\mathcal D}^T)/2, (A_{\mathcal D} -
%% A_{\mathcal D}^T)/2i$ for ${\mathcal D}$ that are not half-turn
%% symmetric.
We begin with a proposition that relates half-turn rotation and matrix transpose.

    \begin{proposition}\label{transpandrotprop}
      Let ${\mathcal D}$ be an oriented $2n$-vertex chord
      diagram, crossing or noncrossing. We have the
      following.
      \begin{equation}\label{transpandroteqn}
        m(\ket{R_{180}{\mathcal D}}) = m(\ket{\mathcal D})^T
      \end{equation}
    \end{proposition}

    \begin{proof}
Let ${\mathcal D}=\{(a_k,b_k)\}_k$ so that $R_{180}{\mathcal
  D}=\{(a_k+n,b_k+n)\}_k$.

Given a $2n$-bit string $K=k_1k_2\ldots k_{2n}$, let $R_{180}K$ denote
the string $R_{180}K = k_{1+n}k_{2+n}\ldots k_{2n+n}$, where
addition in the subscripts is taken mod $2n$. Thus if $K=IJ$ is the
concatenation of $n$-bit strings $I,J$, then $R_{180}K = JI$.

Let $\ket{\mathcal D}=\sum_K
c_K \ket{K}$ be the expansion of $\ket{\mathcal D}$ in the computational basis. If a $2n$-bit
string $K=k_1k_2\ldots k_{2n}$ meets the criterion
$$ k_{a_\ell} = k_{b_\ell}^c, 1\leq \ell\leq n
$$
then the bit string $K'=R_{180}K$ satisfies
$$ k'_{a_\ell} =k_{a_\ell -n} = k_{b_\ell -n}^c = (k'_{b_\ell})^c, 1\leq \ell\leq n.
$$
Using equation~(\ref{diagramstatecoeff}), we have
$$
  \ket{R_{180}{\mathcal D}} = \sum_K c_K \ket{R_{180}K} = \sum_{IJ} c_{IJ} \ket{J}\bra{I} =
  m(\ket{\mathcal D})^T.
  $$
    \end{proof}

The next proposition establishes a detail about the sign in the equation
$\ket{\mathcal D}=\pm\ket{R_{180}{\mathcal D}}$ for diagrams ${\mathcal D}$
with half-turn symmetry. We begin with a Lemma.

      \begin{lemma}\label{midcrossct}
  Let ${\mathcal D}$ be a $2n$-vertex chord diagram, crossing or
  noncrossing. The number of chords that cross the ``midline'', that is,
  the chords that have one vertex in the set $\{1,2,\ldots,n\}$ and the
  other vertex in the set $\{n+1,n+2,\ldots,2n\}$, has the same parity
  as $n$.
      \end{lemma}

      \begin{proof}
Let $c$ be the number of chords that have one vertex in the first half
$\{1,2,\ldots,n\}$ of the vertices and one vertex in the second half
$\{n+1,n+2,\ldots,2n\}$ of the vertices. The number of chords that have
both vertices in the first half must be equal to the number of chords
that have both vertices in the second half, so $n-c$ is even.
      \end{proof}

      \begin{proposition}\label{halfturnsignformandA}
      Suppose that ${\mathcal D}$ has half-turn symmetry, so that $\ket{\mathcal D} =
      \pm\ket{R_{180}{\mathcal D}}$. Then the sign is determined by $n$,
      and we have
      \begin{equation}\label{symmdr180sign}
       \ket{\mathcal D}
      = (-1)^n\ket{R_{180}{\mathcal D}}. 
      \end{equation}
      \end{proposition}

      \begin{proof}
 Let $c$ be the number of chords in ${\mathcal D}$ that join
      vertices in the first half $\{1,2,\ldots,n\}$ with vertices in the
      second half $\{n+1,n+2,\ldots, 2n\}$. By Lemma~\ref{midcrossct},
      $c$ has the same parity as $n$, so we have $(-1)^c=(-1)^n$.
        Each oriented chord $(a,b)$ in
      ${\mathcal D}$ is mapped to $(a+n,b+n)$ in $R_{180}{\mathcal
        D}$. The number of orientation reversals accounts for the global
        sign $(-1)^c=(-1)^n$.
      \end{proof}

      Applying $m$ to both sides of~(\ref{symmdr180sign}), and then
      multiplying both sides by $\Pizza$, we have the following Corollary.
      \begin{corollary}\label{symmdar180sign}
        If ${\mathcal D}$ has half-turn symmetry, then we have the
        following.
        \begin{align}
          m(\ket{\mathcal D}) &= (-1)^n m(\ket{R_{180}{\mathcal D}})\\
       \label{symmdar180signeqn}
          A_{\mathcal D} &= (-1)^n A_{R_{180}{\mathcal D}}
        \end{align}
      \end{corollary}
      
    \begin{proposition}\label{pizzadsymmpair}
            Let ${\mathcal D}$ be an oriented $2n$-vertex chord
      diagram, crossing or noncrossing. We have the
      following.
      \begin{equation}\label{pizzadsymmpaireqn}
        A_{R_{180}{\mathcal D}} = 
      (-1)^n A_{\mathcal D}^T
      \end{equation}
    \end{proposition}
    
    \begin{proof}
      Multiplying both sides of~(\ref{transpandroteqn}) on the left by $\Pizza$, we
      have
      \begin{equation*}
       \Pizza m(\ket{R_{180}{\mathcal D}}) = \Pizza m(\ket{\mathcal D})^T.
      \end{equation*}
      Using Proposition~\ref{pizzaoperatorproplemma}, part~(ii), the
      right side becomes
      \begin{align*}
        (-1)^n\Pizza^T  m(\ket{\mathcal D})^T 
        &= (-1)^n\left[m(\ket{\mathcal D})\Pizza\right]^T\\
        &= (-1)^n\left[\Pizza m(\ket{\mathcal  D})\right]^T
        \hspace*{.2in}\text{(by Prop~\ref{pizzatimesdiagramprop}, part~(i))}.
      \end{align*}
    We conclude that~(\ref{pizzadsymmpaireqn}) holds.
    \end{proof}

    The following Corollary follows immediately from
    Corollary~\ref{symmdar180sign} and Proposition~\ref{pizzadsymmpair}.
    \begin{corollary}\label{symmdtosymmacor}
      If ${\mathcal D}$ has half-turn symmetry, then $A_{\mathcal
        D}=A_{\mathcal D}^T$.
    \end{corollary}
    
    Now we construct a set of Hermitian matrices. For every
    ${\mathcal D}$ in $\ncc$, let $S_{\mathcal D}$ denote the set
    $$S_{\mathcal D}=\left\{\frac{A_{\mathcal D}+A_{\mathcal D}^T}{2}, \frac{A_{\mathcal
          D}-A_{\mathcal D}^T}{2i}\right\}.
    $$
    Because the $A_{\mathcal D}$ have real entries, the matrices in
    $S_{\mathcal D}$ are Hermitian. To extract a basis from the
    collection $\bigcup_{{\mathcal D}\in \ncc} S_{\mathcal D}$, 
    we need to weed out linear dependencies that
    arise from the fact that $A_{\mathcal D}^T = A_{R_{180}\mathcal D}$ (Proposition~\ref{pizzadsymmpair}).
    We categorize diagrams in $\ncc$ into two types, corresponding to
    when the underlying unoriented chord diagram either does have or
    does not have half-turn rotational symmetry.
    \begin{align*}
      \ncc_{\text{symm}} &= \{{\mathcal D}\in \ncc \colon \ket{\mathcal
        D}=\pm \ket{R_{180}{\mathcal D}}  \}\\
      \ncc_{\text{nonrot}} &= \left\{
    {\mathcal D}\in \ncc \colon \ket{\mathcal D} \neq \pm \ket{R_{180}{\mathcal D}}
    \right\}
    \end{align*}
    The figures in the left column of Table~\ref{tab:n3eg} show examples of each of these types.
    
    If ${\mathcal D}\in \ncc_{\text{symm}}$, then $A_{\mathcal D}$ is
    real symmetric by Corollary~\ref{symmdtosymmacor}, so
    $S_{\mathcal D}= \{A_{\mathcal D},0\}$. If ${\mathcal D}\in
    \ncc_{\text{nonrot}}$, then $S_{\mathcal D}$ is a set of two nonzero
    Hermitian matrices. But we have redundancies: let ${\mathcal E}$ be
    the {\em unoriented} noncrossing chord diagram with the same diagram
    as ${R_{180}{\mathcal D}}$, and give ${\mathcal E}$ the standard
    orientation (so that the chord $\{a,b\}$ is oriented $(a,b)$ with
    $a<b$) so that ${\mathcal E}$ is in $\ncc$. Then the vectors in the
    set $S_{\mathcal E}$ are the same, up to sign, as the vectors in
    $S_{\mathcal D}$, so that the sets $S_{\mathcal D},S_{\mathcal E}$
    have the same linear span. To eliminate these redundancies, let $R$ be a set
    consisting of a choice of one of the two $\ncc_{\text{nonrot}}$
    diagrams ${\mathcal D},{\mathcal E}$ for each pair of the type just
    described. The choice can be arbitrary, but here is one way to
    construct $R$ explicitly. Write ${\mathcal D}$ in $\ncc$ as a string
    of indices $a_1,b_1,a_2,b_2,\ldots,a_n,b_n$, where $\{a_k,b_k\}$ are
    the unoriented chords in ${\mathcal D}$ with $a_k<b_k$ for all $k$,
    and $a_1<a_2<\cdots <a_n$. Then write ${\mathcal D}<{\mathcal D}'$
    to indicate that ${\mathcal D}$ comes before ${\mathcal D}'$ in
    lexicographical order. Now we can define the set $R$ by
    $$ R =\{{\mathcal D}\in \ncc_{\text{nonrot}}\colon {\mathcal D}<R_{180}{\mathcal D}\}.
    $$
    Now we assemble carefully chosen elements from the sets $S_{\mathcal
      D}$. Let $B$ be the set
    \begin{equation}\label{nccbasissetdef}
      B = \left\{ A_{\mathcal D}\colon {\mathcal D}\in
      \ncc_{\text{symm}} \right\} \cup
\left\{  \frac{A_{\mathcal D}+A_{\mathcal D}^T}{2}, \frac{A_{\mathcal
          D}-A_{\mathcal D}^T}{2i}\colon {\mathcal D}\in R
      \right\}.
    \end{equation}
    To complete the argument that $B$ satisfies the requirements for our
    basis construction, we count dimensions. The dimension of
    $\purewernerspace{2n}$ is the Catalan number $C_n=\frac{1}{n+1}{2n
      \choose n}$~\cite{su2blockstates}. The set $\{\ket{\mathcal D}\colon
    {\mathcal D}\in \ncc\}$ is a $\C$-basis for $\purewernerspace{2n}$,
    so the cardinality of the set $\ncc$ is $C_n$. The
    dimension of the space $\hermitianwernerspace{n}$ is also the
    Catalan number $C_n$~\cite{wernerstructure}. The real linear
    transformation $\purewernerspace{2n}\to \hermitianwernerspace{n}$
    taking $\ket{\mathcal D}$ to $A_{\mathcal D}$ is nonsingular. For
    ${\mathcal D}\in R$, we have $A_{\mathcal D}^T = \pm
    A_{R_{180}{\mathcal D}}$, and the transformation determined by
    $$\left(A_{\mathcal D},A_{\mathcal D}^T\right) \to \left(\frac{A_{\mathcal
        D}+A_{\mathcal D}^T}{2}, \frac{A_{\mathcal D}-A_{\mathcal
        D}^T}{2i}\right)
    $$ is invertible. Thus, the set $B$ is a set of independent Hermitian
    operators with cardinality $C_n$. We conclude that $B$ is a basis
    for $\hermitianwernerspace{n}$. This completes the basis
    construction. We record the result with the following
    Theorem. Table~\ref{tab:n3eg}
    %in Appendix~\ref{nccn3egappdx}
    shows details for the example $n=3$.
    
    \begin{theorem}\label{nccmixedbasisthm}
      {\bf (A basis for $n$-qubit Werner mixed states)} The set $B$
      (given by~(\ref{nccbasissetdef}) above) is a basis for the real
      vector space $\hermitianwernerspace{n}$.  Any $n$-qubit Werner
      invariant density matrix is a unique $\R$-linear combination of
      elements in this basis.
    \end{theorem}

\begin{table}
\spacer\hspace*{-.2in}
}\\ \spacer}
  & \shortstack{(none)\\ \rule{0in}{.5in}}\\ \hline  
\end{tabular}
  \caption{Bases for Werner invariant spaces for $n=3$. The far left
    column shows the five 6-vertex noncrossing chord
    diagrams. The diagrams in the first
    three rows have half-turn rotational symmetry, and the diagrams in
    the last two rows do not (they are half-turn rotations of one
    another). The column with
    the heading `$\ket{\mathcal D}$' is the $\ncc$ basis for the complex 5-dimensional space
    of Werner invariant vectors in the Hilbert space for 6 qubits,
    omitting the normalizing factor $1/\sqrt{8}$. For space efficiency and readability,
  matrices in the column with the heading `$A_{\mathcal D}$' are typeset without brackets or
  parentheses, and the symbols `$+$', `$-$' are used to denote the
  entries $1/\sqrt{8}$, $-1/\sqrt{8}$, respectively. The
    column on the far right is a basis for the real 5-dimensional space
    of Werner invariant Hermitian operators on the Hilbert space for 3
    qubits.  The set $R$ consists of the single diagram ${\mathcal
      D}=\{(1,2),(3,4),(5,6)\}$. }
  \label{tab:n3eg}
\end{table}

\section{Polygon States: 2-party Reduced Density Matrices}\label{rdmentsect}

In this section we show that any 2-party reduced density matrix of the
Werner state $\rho_m$ (equation~(\ref{wernergondef})) is separable for
$m\geq 3$.

Because partial trace commutes with the local unitary action, Werner
invariance of a mixed state is inherited by all of its reduced density
matrices. In particular, any 2-qubit reduced density matrix of a Werner
state is also a Werner state, which can be written in the form
$$ \rho = \lambda \frac{\Id}{4} + (1-\lambda)\ket{s}\bra{s}
$$ for some $0\leq \lambda \leq 4/3$. The state $\rho$ is entangled if
$\rho_{00,00}<1/6$ and $\rho$ is separable if $\rho_{00,00}\geq
1/6$~\cite{werner89,unanyan2007decomposition}.

Choose two qubits $a,b$, $1\leq a<b\leq m$, and let $\rho$ be the
2-qubit Werner state
$$\rho =\rho_m^{a,b} =   \tr_{\text{(all but $a,b$)}}\rho_m.
$$ It will be convenient to use the following labels (see
Section~\ref{prelimsect} for the definition of aperiodic $m$-bit
string).
\begin{align*}
  A(m) &= \text{ number of aperiodic $m$-bit strings}\\
  P(m) &= \text{ number of periodic $m$-bit strings}\\
  A_{00}(m) &= \text{ number of aperiodic $m$-bit strings with 00 in $a,b$}\\  
  P_{00}(n) &= \text{ number of periodic $m$-bit strings with 00 in $a,b$}  
\end{align*}
We shall make use of the following elementary relationships.
\begin{align}\label{totaldegenandnondegen}
  2^m &= A(m) + P(m)\\
  2^{m-2} &= A_{00}(m) + P_{00}(m)\\  
  P_{00}(m) &\leq P(m)\\
  P(m) &= \sum_{d|m,d<m} A(d)\\ \label{sumofgeomcompare}
  P(m) &\leq \sum_{i=1}^{\lfloor m/2\rfloor} A(i) \leq
  \sum_{i=0}^{\lfloor m/2\rfloor} 2^i = 2^{\lfloor m/2\rfloor+1}-1
\end{align}
The left-most inequality in~(\ref{sumofgeomcompare}) comes from the fact
that the number of divisors of $m$ is upper-bounded by $\lfloor
m/2\rfloor$.

We begin by obtaining an expression for $\rho_{00,00}$. Suppose an
$m$-bit string $I$ is aperiodic, with
$i_a=i_b=0$. From the definition~(\ref{cyclicistatedef}) for $C(I)$, we
have
\begin{equation}\label{cicidagg}
   C(I)C(I)^\dagger = \frac{1}{m} \sum_{k,\ell=0}^{m-1}
\omega^{k-\ell}\ket{\pi^k I}\bra{\pi^\ell I}.
\end{equation}
The only term in the sum on the right side that gives a nonzero
contribution to the partial trace over all qubits but $a,b$ is for
$k=\ell=0$, so we have
\begin{equation}\label{trcicidagg}
   \bra{0_a0_b} \tr_{\text{(all but $a,b$)}} C(I)C(I)^\dagger
\ket{0_a0_b} = \frac{1}{m}.
\end{equation}
From~(\ref{cicidagg}), it is easy to see that 
$C(I)C(I)^\dagger = C(\pi^k I)C(\pi^k I)^\dagger$ for $0\leq k\leq
m-1$, so~(\ref{trcicidagg}) becomes
\begin{equation}\label{trsumcicidagg}
   \bra{0_a0_b} \tr_{\text{(all but $a,b$)}} \sum_{k=0}^{m-1}C(\pi^k
I)C(\pi^k I)^\dagger
\ket{0_a0_b} = 1.
\end{equation}
From definition~(\ref{wernergondef}), it follows that
\begin{equation}\label{r0000expr}
    \rho_{00,00} = \frac{A_{00}(m)}{A(m)}.
\end{equation}
Applying~(\ref{totaldegenandnondegen})--(\ref{sumofgeomcompare}), we
have
\begin{align}
  \rho_{00,00} &=
  \frac{A_{00}(m)}{A(m)}\\
  &=\frac{2^{m-2}-P_{00}(m)}{2^m-P(m)}\\
  &\geq \frac{2^{m-2}-P(m)}{2^m}\\
  &\geq   \frac{2^{m-2}-2^{\lfloor m/2\rfloor +1}+1}{2^m}\\
  &\geq   \frac{2^{m-2}-2^{\lfloor m/2\rfloor +1}}{2^m}\\  
  &= \frac{1}{4} - 2^{\lfloor m/2\rfloor +1-m}\\
    \label{rho00lowerbd}
  &= 
  \begin{cases}
    \frac{1}{4} - 2^{(2-m)/2} & \text{ $m$ even}\\
    \frac{1}{4} - 2^{(1-m)/2} & \text{ $m$ odd}
  \end{cases}.
\end{align}
It is clear that~(\ref{rho00lowerbd}) increases as $m$ increases. It is
easy to check that~(\ref{rho00lowerbd}) is equal to $3/16>1/6$ for
$m=9,10$, so therefore~(\ref{rho00lowerbd}) is larger than $1/6$ for
$m\geq 9$. Table~\ref{tab:2particleent}
%in Appendix~\ref{rdmentrytableappdx}
shows that $\rho^{a,b}_{00,00}\geq 1/6$ for all possibilities for $a,b$,
for $3\leq m\leq 8$. We record the result of this section as the
following Theorem.

\begin{theorem}
  Let $\rho_m$ denote the $m$-qubit mixed Werner
  state~(\ref{wernergondef}) for some $m\geq 3$. Let $a,b$ be any two
  qubits $1\leq a,b\leq m$, and let $\rho=\rho_m^{a,b}$ be the 2-qubit
  reduced density matrix of $\rho_m$ is the subsystem consisting of
  qubits $a,b$. Then $\rho$ is separable.
\end{theorem}

  \begin{table}
    \begin{center}
    \begin{tabular}{ccr}
      $m$ & $|a-b|$ & $\rho_{00,00}\rule{.2in}{0in}$ \\ \hline
3	&	1	&	$	1/6	\approx	0.1667	$	\\
4	&	1	&	$	1/4	\approx	0.2500	$	\\
4	&	2	&	$	1/6	\approx	0.1667	$	\\
5	&	1	&	$	7/30	\approx	0.2333	$	\\
5	&	2	&	$	7/30	\approx	0.2333	$	\\
6	&	1	&	$	7/27	\approx	0.2593	$	\\
6	&	2	&	$	13/54	\approx	0.2407	$	\\
6	&	3	&	$	2/9	\approx	0.2222	$	\\
7	&	1	&	$	31/126	\approx	0.2540	$	\\
7	&	2	&	$	31/126	\approx	0.2460	$	\\
7	&	3	&	$	31/126	\approx	0.2460	$	\\
8	&	1	&	$	1/4	\approx	0.2500	$	\\
8	&	2	&	$	1/4	\approx	0.2500	$	\\
8	&	3	&	$	1/4	\approx	0.2500	$	\\
8	&	4	&	$	7/30	\approx	0.2333	$	\\
    \end{tabular}
    \end{center}
    \caption{Values of $\rho_{00,00}$ for $3\leq m\leq 8$ showing
      separability (any value $\geq 1/6$ implies $\rho_{00,00}$ is
      separable). Table values depend only on the distance $|a-b|$
      because of the cyclic symmetry of $\rho_m$.}
    \label{tab:2particleent}
  \end{table}

\section{Outlook}

  The eventual goal for constructing a basis for mixed
  Werner states is to characterize entanglement properties and
  identify resource states in terms of coefficients with respect to that
  basis. A first step will be to determine constraints on coefficients
  that correspond to {\em states}, i.e., operators that are positive
  semidefinite and have trace 1. {There will be two
    immediately interesting questions: in what ways can we use the
    bases constructed in Section~\ref{nccbasissection} for $n\geq 3$
    qubits to generalize or extend Werner and Eggeling's basis
   in~\cite{eggelingwernerPhysRevA.63.042111}? Second, can we generalize
   and extend the basis construction to qudits?}

  The separability result in Section~\ref{rdmentsect} for 2-qubit reduced density matrices of Werner
  states $\rho_m$ provides motivation to seek further results in
  characterizing separability properties for mixtures of polygon diagram
  states, constructed in~\cite{wernerstructure}. We hope to identify
  distributed entanglement protocols that will exploit these states.

  \subsection*{Acknowledgments} This work was supported by Lebanon
  Valley College and by grant \#PHY-2011074 from the National Science
  Foundation. 

%\section{References}

    \appendix

    \section{Proofs of Propositions}

    {\bf Proof Werner invariance criteria.}
    Formulas~(\ref{Acondpure})--(\ref{Ccondmixed}) are special cases of
    more general formulas for the action of the Lie algebra of the local
    unitary group on pure and mixed states. {(For example,
    see~\cite{symmmixed}. See~\cite{guhne2009entanglement} for a
    connection with angular momentum.)} For the sake of self-containedness, here is
    a proof.

  The Lie algebra
  $L(SU(2))$ of the special unitary group $SU(2)$ is the real vector
  space of $2\times 2$ skew-Hermitian matrices with trace zero, and is
  generated (by real linear combinations and the bracket operation) by
  the operators $iZ,iX$. Given a group action $\Phi\colon SU(2)\times V \to V$ on a
  vector space $V$, there is a Lie algebra action $L(\Phi)\colon L(SU(2))\times V \to V$ 
  on $V$ given by by $L(\Phi)(M) v = \left.\frac{d}{dt}\right|_{t=0} \Phi(\exp(tM)) v$, where $M =
  i(aX+bY+cZ)$ for some real coefficients $a,b,c$, and $Y$ is the Pauli
  $Y$ operator. If both generators $iZ,iX$ annihilate a vector $v$ in
  $V$, then $v$ is fixed by $\exp(it(aX+bY+cZ)$ for all real
  $t$. To obtain the results in the Lemma, we apply this basic
  observation to the actions of $SU(2)$ on 
  $\purestatespace{m}$ and
  $\matrixspace{m}$ by the standard local unitary actions
  \begin{align}\label{suwerneractionpure}
    \Phi_{\text{pure}}(U) \ket{\psi} &= U^{\otimes m}\ket{\psi}\\ \label{suwerneractionmixed}
    \Phi_{\text{mixed}}(U) \rho &= U^{\otimes m}\rho (U^\dagger)^{\otimes m}
  \end{align}
  The Werner invariance conditions in the Lemma arise by taking
  derivatives on the right sides of the following equations. 
  \begin{align*}
0&= L(\Phi_{\text{pure}})(iZ)\ket{\psi} =  \left.\frac{d}{dt}\right|_{t=0} \exp(itZ)^{\otimes m}\ket{\psi}\\
0&= L(\Phi_{\text{pure}})(iX)\ket{\psi} =  \left.\frac{d}{dt}\right|_{t=0} \exp(itX)^{\otimes m}\ket{\psi}\\
0&= L(\Phi_{\text{mixed}})(iZ)\rho =  \left.\frac{d}{dt}\right|_{t=0} \exp(itZ)^{\otimes m}\rho\exp(-itZ)^{\otimes m}\\
0&= L(\Phi_{\text{mixed}})(iX)\rho =  \left.\frac{d}{dt}\right|_{t=0} \exp(itX)^{\otimes m}\rho\exp(-itX)^{\otimes m}
  \end{align*}
  Finally, we observe that if a pure state $\ket{\psi}$ is fixed by
  every $U$ in $SU(2)$ acting by~(\ref{suwerneractionpure}), then
  $\ket{\psi}$ is fixed, up to a phase factor, by any $V$ in $U(2)$,
  since any particular $V\in U(2)$ can be written $e^{i\theta}U$ for some real $\theta$ and
  some $U\in SU(2)$. (No such phase adjustment is necessary
  for~(\ref{suwerneractionmixed}).) This concludes the proof.

    The following Lemma gives computationally useful forms for
    the Werner invariance criteria~(\ref{Acondpure})--(\ref{Ccondmixed}). The proof is straightforward
    checking. For more general formulas for which these are special
    cases, see~\cite{symmmixed}.
    \begin{lemma}
      {\bf (Detailed forms of Werner invariance criteria)}
    Let $\ket{\psi}=\sum_I c_I \ket{I}$ and let $\rho = \sum_{I,J}
    \rho_{I,J} \ket{I}\bra{J}$ be a pure state and a mixed state,
    respectively, of $m$-qubits, with respect to the computational
    basis. The following hold.
    \begin{align}
    \left(\sum_k Z^{(k)}\right)\cdot \ket{\psi} &= \sum_I \left( \sum_k
    (-1)^k 
    \right)c_{I} \ket{I}\label{Acondpuredetail}\\
    \left(\sum_k X^{(k)}\right)\cdot \ket{\psi} &= \sum_I
    \left(\sum_{k} c_{I_k}\right) \ket{I} \label{Ccondpuredetail}\\
    \left[\left(\sum_k  Z^{(k)}\right),\rho\right] &= \sum_{IJ}
    \rho_{IJ} \left(\sum_{k\colon i_k\neq j_k}
    (-1)^{i_k}\right)\ket{I}\bra{J}\label{Acondmixeddetail}\\
    \left[\left(\sum_k X^{(k)}\right),\rho\right] &=
    \sum_{IJ}\left(\sum_{k,\ell = 1}^n (\rho_{I_k,J} -
    \rho_{I,J_\ell})\right) \ket{I}\bra{J}     \label{Ccondmixeddetail}
  \end{align}
    \end{lemma}

    {\bf Proof of Proposition~\ref{pizzatimesdiagramprop}.}
  Let
  $\ket{\mathcal D} = \sum_{K,J}c_{KJ} \ket{K}\ket{J}$.
  Because $\ket{\mathcal D}$ is a product of singlets, we
  have
  \begin{align} \label{idxwtcomplKJ}
     n &= \wt K + \wt J \hspace*{.2in}\text{ and }\\
\label{coeffidxcompl}
   c_{K^cJ^c} &= (-1)^n c_{KJ} 
  \end{align}
  for all $K,J$ such that $c_{KJ}\neq 0$.
  From~(\ref{idxwtcomplKJ}) we  have
  \begin{equation}
(-1)^{\wt K}=(-1)^n(-1)^{\wt J} \label{idxwtcomplexp}
  \end{equation}
  for all $K,J$ such that $c_{KJ}\neq 0$. As a special case
  of~(\ref{idxwtcomplKJ}), we have
    \begin{equation}\label{idxwtcomplI}
   n= \wt I + \wt I^c 
  \end{equation}
    for all $I$
in the expression $\Pizza=\sum_I (-1)^{\wt I} \ket{I}\bra{I^c}$ (part~(i) of
Proposition~\ref{pizzaoperatorproplemma}).

(Proof of statement~(i)) We have
  \begin{align} \label{firstprod}
    \Pizza m(\ket{\mathcal D})
    &= \left(\sum_I (-1)^{\wt I} \ket{I}\bra{I^c}\right)
    \left(\sum_{K,J} c_{KJ} \ket{K}\bra{J}\right) \\ \label{firstprodsimp}
    &= \sum_{I,J} (-1)^{\wt I} c_{I^cJ}\ket{I}\bra{J}\\ \nonumber
    &= \sum_{K,J} (-1)^{\wt K} c_{K^cJ}\ket{K}\bra{J}
    \hspace*{.2in}\text{(substitute  $I\leftrightarrow K$)}\\ \nonumber
    &= \sum_{K,J} (-1)^n(-1)^{\wt K}(-1)^n c_{K^cJ^c}\ket{K}\bra{J^c}
    \hspace*{.2in}\text{(substitute  $J\leftrightarrow J^c$)}\\ \nonumber
    &= \sum_{K,J} (-1)^{\wt J} c_{KJ}\ket{K}\bra{J^c}
    \hspace*{.2in}\text{(using (\ref{coeffidxcompl}) and
      (\ref{idxwtcomplexp}))}\\ \nonumber
    &= \left(\sum_{K,J} c_{KJ} \ket{K}\bra{J}\right)\left(\sum_I
    (-1)^{\wt I} \ket{I}\bra{I^c}\right) \\ \nonumber
    &=    m(\ket{\mathcal D}) \Pizza.
  \end{align}

  (Proof of statement~(ii))  To show Werner invariance of
  $\Pizza m(\ket{\mathcal D})$, we check that conditions~(\ref{Acondmixeddetail}),
  (\ref{Ccondmixeddetail}) hold for the expression~(\ref{firstprodsimp}). 
  Because~(\ref{idxwtcomplKJ}) and~(\ref{idxwtcomplI}) are satisfied in~(\ref{firstprod}), we
  must have
  $$n = \wt I + \wt J
  $$
  in~(\ref{firstprodsimp}) for every $\ket{I}\bra{J}$ term with nonzero
  coefficient. From this it follows that 
  \begin{equation}
    \sum_{k\colon i_k\neq j_k} (-1)^{i_k}=0
  \end{equation}
  for all $I,J$ such that $\ket{I}\bra{J}$ appears with nonzero
  coefficient in~(\ref{firstprodsimp}), and
  therefore~(\ref{Acondmixeddetail}) is zero.

  For~(\ref{Ccondmixeddetail}), we have
  \begin{align}
        \sum_{k,\ell = 1}^n (\rho_{I_k,J} - \rho_{I,J_\ell}) &=
        \sum_{k,\ell} ((-1)^{\wt I_k} c_{{I_k}^cJ} - (-1)^{\wt I}
        c_{I^cJ_\ell})\\
        &= (-1)^{(\wt I +1)}\sum_{k,\ell}(c_{I_k^cJ} + c_{I^cJ_\ell})
  \end{align}
  The last expression is (a sign times) the coefficient of $\ket{I^cJ}$
  in the expansion of
  $\left(\sum_{k=1}^{2n} C^{(k)}\right) \ket{\mathcal D})$, and so this
  quantity is zero (by~(\ref{Ccondmixeddetail})) because $\ket{\mathcal
    D}$ is a product of singlets. We conclude
  that~(\ref{Ccondmixeddetail}) is zero for all $I,J$.

%  \section{Mixed Werner basis construction details for $n=3$}\label{nccn3egappdx}

%\section{2-party Reduced Density Matrix entries for $n\leq 8$}\label{rdmentrytableappdx}

\end{document}